\documentclass{article}

\usepackage[final]{neurips_ai4d3_2023}

\usepackage[utf8]{inputenc} 
\usepackage[T1]{fontenc}    
\usepackage{hyperref}       
\usepackage{url}            
\usepackage{booktabs}       
\usepackage{amsfonts}       
\usepackage{nicefrac}       
\usepackage{microtype}      
\usepackage{xcolor}         
\usepackage{adjustbox}
\usepackage{amsmath} 
\usepackage[flushleft]{threeparttable}
\usepackage{multirow}
\usepackage{graphicx}
\usepackage{subcaption}

\usepackage{amsmath}
\usepackage{bm}

\def\eqref#1{equation~\ref{#1}}

\def\1{\bm{1}}

\def\vx{{\bm{x}}}

\DeclareMathAlphabet{\mathsfit}{\encodingdefault}{\sfdefault}{m}{sl}
\SetMathAlphabet{\mathsfit}{bold}{\encodingdefault}{\sfdefault}{bx}{n}

\def\gG{{\mathcal{G}}}

\def\gL{{\mathcal{L}}}

\def\gR{{\mathcal{R}}}
\def\gS{{\mathcal{S}}}

\def\gX{{\mathcal{X}}}

\newcommand{\R}{\mathbb{R}}

\let\oldAA\AA
\renewcommand{\AA}{\text{\normalfont\oldAA}}
\usepackage{cancel}
\usepackage{xcolor}
\usepackage{tabularray}
\title{PDB-Struct: A Comprehensive Benchmark for Structure-based Protein Design}

\newcommand{\revise}[1]{#1}
\author {
    Chuanrui Wang\textsuperscript{\rm 1,2} \quad
    Bozitao Zhong \textsuperscript{\rm 1,2} \quad
    Zuobai Zhang\textsuperscript{\rm 1,2} \quad \\
    \textbf{Narendra Chaudhary}\textsuperscript{\rm 3} \quad
    \textbf{Sanchit Misra}\textsuperscript{\rm 3} \quad
    \textbf{Jian Tang}\textsuperscript{\rm 1,4,5$\,\dagger$} \\
    \textsuperscript{\rm 1} Mila - Qu\'ebec AI Institute \quad
    \textsuperscript{\rm 2} Universit\'e de Montr\'eal \quad
    \textsuperscript{\rm 3} Intel Parallel Computing Lab \quad \\
    \textsuperscript{\rm 4} HEC Montr\'eal \quad
    \textsuperscript{\rm 5} CIFAR AI Chair \quad \\
    \texttt{\{chuanrui.wang, bozitao.zhong, zuobai.zhang\}@mila.quebec}, \\
    \texttt{\{narendra.chaudhary, sanchit.misra\}@intel.com},
    \texttt{jian.tang@hec.ca}
}

\begin{document}

\maketitle

\begin{abstract}

Structure-based protein design has attracted increasing interest, with numerous methods being introduced in recent years.
However, a universally accepted method for evaluation has not been established, since the wet-lab validation can be overly time-consuming for the development of new algorithms, and the \textit{in silico} validation with recovery and perplexity metrics is efficient but may not precisely reflect true foldability.
To address this gap, we introduce two novel metrics: refoldability-based metric, which leverages high-accuracy protein structure prediction models as a proxy for wet lab experiments, and stability-based metric, which assesses whether models can assign high likelihoods to experimentally stable proteins.
We curate datasets from high-quality CATH protein data, \revise{high-throughput \textit{de novo} designed proteins, and mega-scale experimental mutagenesis experiments},
and in doing so, present the \textbf{PDB-Struct} benchmark that evaluates both recent and previously uncompared protein design methods.
Experimental results indicate that ByProt, ProteinMPNN, and ESM-IF perform exceptionally well on our benchmark, while ESM-Design and AF-Design fall short on the refoldability metric.
We also show that while some methods exhibit high sequence recovery, they do not perform as well on our new benchmark.
Our proposed benchmark paves the way for a fair and comprehensive evaluation of protein design methods in the future. 
\revise{Code is available at \href{https://github.com/WANG-CR/PDB-Struct}{https://github.com/WANG-CR/PDB-Struct}.}

\end{abstract}
\section{Introduction}

Designing new proteins with desired properties is a crucial task in bioengineering~\citep{huang2016coming}. 
It aids in developing therapies, crafting novel antibodies, and exploring the uncharted realm of proteins beyond those found in nature.
Structure-based protein design has emerged as the predominant approach for de novo protein design, owing to its versatile application across proteins with well-defined structures. In recent years, the integration of deep learning has enhanced the capabilities of structure-based protein design, yielding notable results~\citep{ingraham2019generative, jing2020learning, dauparas2022robust, zheng2023structure}. This progress is evidenced by the creation of miniproteins specifically engineered to bind particular targets~\citep{cao2022design}, enzymes fine-tuned for new substrates~\citep{yeh2023novo}, and innovative antibodies~\citep{luo2022antigen,shi2022protein}.


Benchmarking these methods is of paramount importance; 
however, 
current benchmarks have limitations. 
While experimental validation of generated sequences is costly~\citep{ladd1977structure, bai2015cryo, dauparas2022robust}, the \textit{in silico} proxy, which calculates sequence recovery and perplexity~\citep{jing2020learning} on a test set of natural proteins, serves as an effective replacement but may not accurately reflect real-world foldability.
The recovery is calculated as the similarity between the designed sequence and the ground truth sequence, but high sequence similarity can not sufficiently imply the ability to fold into similar structures because even single mutations could cause a protein to misfold, such as Alzheimer's and cystic fibrosis~\citep{cohen2003therapeutic,qu1997cystic}. 
The perplexity evaluates the uncertainty of a model's predictions by measuring the likelihood that the protein design model assigns to the ground truth sequence. 
However, the ground truth sequence only provide a point probability mass function instead of the true distribution landscape.
Furthurmore, protein design methods compute pseudo-likelihoods based on varying assumptions, making these scores incomparable. For instance, one-shot prediction models~\citep{gao2023pifold} compute it under the conditional independence assumption, whereas autoregressive models~\citep{ingraham2019generative} do not use this assumption.
To address these limitations and complement the known metrics, we propose two novel metrics to benchmark structure-based protein design methods.
The first metric, termed "refoldability", assesses the quality of designed sequences.
This quality is determined by two factors: whether the designed sequences can fold into a stable structure and whether this stable structure is similar to the input structure.
We utilize the atom-level protein structure prediction models~\citep{jumper2021highly,mirdita2022colabfold,lin2022language,wu2022high} to predict structures for the designed sequences, and calculate the TM score between the predicted structure and the input structure, as well as the pLDDT score reflecting the folding stability.
The second metric, termed "stability-based metric", 
measures whether the 
the protein design methods can accurately estimate protein sequence landscape. 
We have curated datasets comprising one structure template and multiple sequences, derived from high-throughput \textit{de novo} protein design and mutagenesis experiments~\citep{rocklin2017global, tsuboyama2023mega}. 
Within these sequences, better protein design methods should assign higher probability to the sequences with higher experimental stability score, returning a higher stability-based Spearman correlation.
Using these metrics and datasets, we evaluate the latest models as well as some previously unexamined ones. We have named this benchmark \textbf{PDB-Struct}, 
signifying a \textbf{struct}ure-based \textbf{p}rotein \textbf{d}esign \textbf{b}enchmark. 
Our contributions can be outlined as follows:
\begin{itemize}
    \item We introduced two evaluation metrics based on carefully curated datasets.
    \item We ran experiments with popular structure-based protein design models and have established a fresh benchmark termed \textbf{PDB-Struct}.
    \item The proposed benchmark is the first that compares encoder-decoder based protein design methods together with structure-prediction based methods.
    \item By analysing the benchmark results, we outline the pros and cons of each protein design model, providing guidance for protein scientists when choosing a model.
\end{itemize}

\section{Preliminaries}
\subsection{Problem Definition}
Protein can be represented as a pair of amino acid sequence and structure $(\gS,\gX)$, where $\gS=[s_1,s_2,\cdots,s_{n}]$ denotes its sequence of $n$ residues with $s_i\in\{1,...,20\}$ indicating the type of the $i$-th residue, and $\gX=[\vx_1,\vx_2...,\vx_{n}]\in\R^{n\times 4 \times 3}$ denotes its structure with $\vx_i$ representing the Cartesian coordinates of the $i$-th residue's backbone atoms, including N, C-$\alpha$, C and O.
The challenge posed by the structure-based protein design is to elucidate an effective model $\theta$ capable of learning the underlying mapping from the provided structure data to the corresponding sequence distribution, and then generate novel sequences $\hat{\gS} \sim p_{\theta}(\gS \vert \gX)$.


\subsection{Existing Methods}

\paragraph{Encoder-Decoder Model}
Traditional methods encode 3D structure data using hand-crafted features or direct atom positions, typically employing MLPs~\citep{o2018spin2,li2014direct} and CNNs~\citep{qi2020densecpd, anand2022protein}. Alternatively, viewing protein structure as a k-NN graph of amino acids retains spatial information, making GNNs a favored encoder.
StructTrans employ graph-based self-attention modules in their encoder-decoder framework and decode in an autoregressive manner\citep{ingraham2019generative}.
Further advancements have been made by GVP~\citep{jing2020learning}, ProteinMPNN~\citep{dauparas2022robust} and ESM-Inverse Folding~\citep{hsu2022learning}, both showcasing significant improvements. 
Some of the latest models suggest decoding residues conditionally independently given their structure, which accelerates the generation process without compromising sequence recovery\citep{gao2023pifold}.
Moreover, inspired by the achievements in protein language modeling, ByProt introduced a structure adapter to incorporate ESM2 model\citep{lin2022language}, then decode as iterative refinement and boasts high sequence recovery\citep{zheng2023structure}.
Works on graph-based encoder-decoder paradigms are emerging\citep{tan2022generative,gao2023knowledge,mao2023modeling}, setting new benchmark in sequence recovery metric.


\paragraph{Structure Prediction based Model}
Models of this kind utilize pretrained structure prediction models or pretrained language models~\citep{yang2020improved, jumper2021highly} to compute an energy function, and then utilise different sampling strategies to generate samples.
\cite{wang2022scaffolding} proposed to sample with thousands of gradient steps. Alternatively, \cite{verkuil2022language} proposed to perform Markov chain Monte Carlo sampling steps combined with simulated annealing, all to minimize the loss functions defined by protein structure prediction models and the structure condition. 
Similarly, hallucination methods aim to maximize the KL divergence between the predicted structures and a background distribution~\citep{anishchenko2021novo, hie2022high}. 
However, it should be noted that the sampling process in these models tends to be slower than that in encoder-decoder models.

\paragraph{Diffusion-based Model}
Diffusion models~\citep{ho2020denoising} offer an alternative to generate samples through denoising, and they potentially offer advantages when learning from limited data~\citep{zaidi2022pre}.
\cite{yi2023graph} performs denoising in the graph attribute space and achieves high sequence recovery.
There are other models applying diffusion models in the discrete sequence space, such as EvoDiff~\citep{alamdari2023protein} and, ProteinGenerator~\citep{lisanza2023joint}.
Other works, like Chroma~\citep{ingraham2022illuminating} and RFDiffusion~\citep{watson2022broadly}, apply denoising in the structure space.
Since they have not released the code, or do not apply to structure-based protein sequence design, we are not evaluating them at the moment.


\begin{figure}[ht]
\centering
\includegraphics[width=0.8\textwidth]{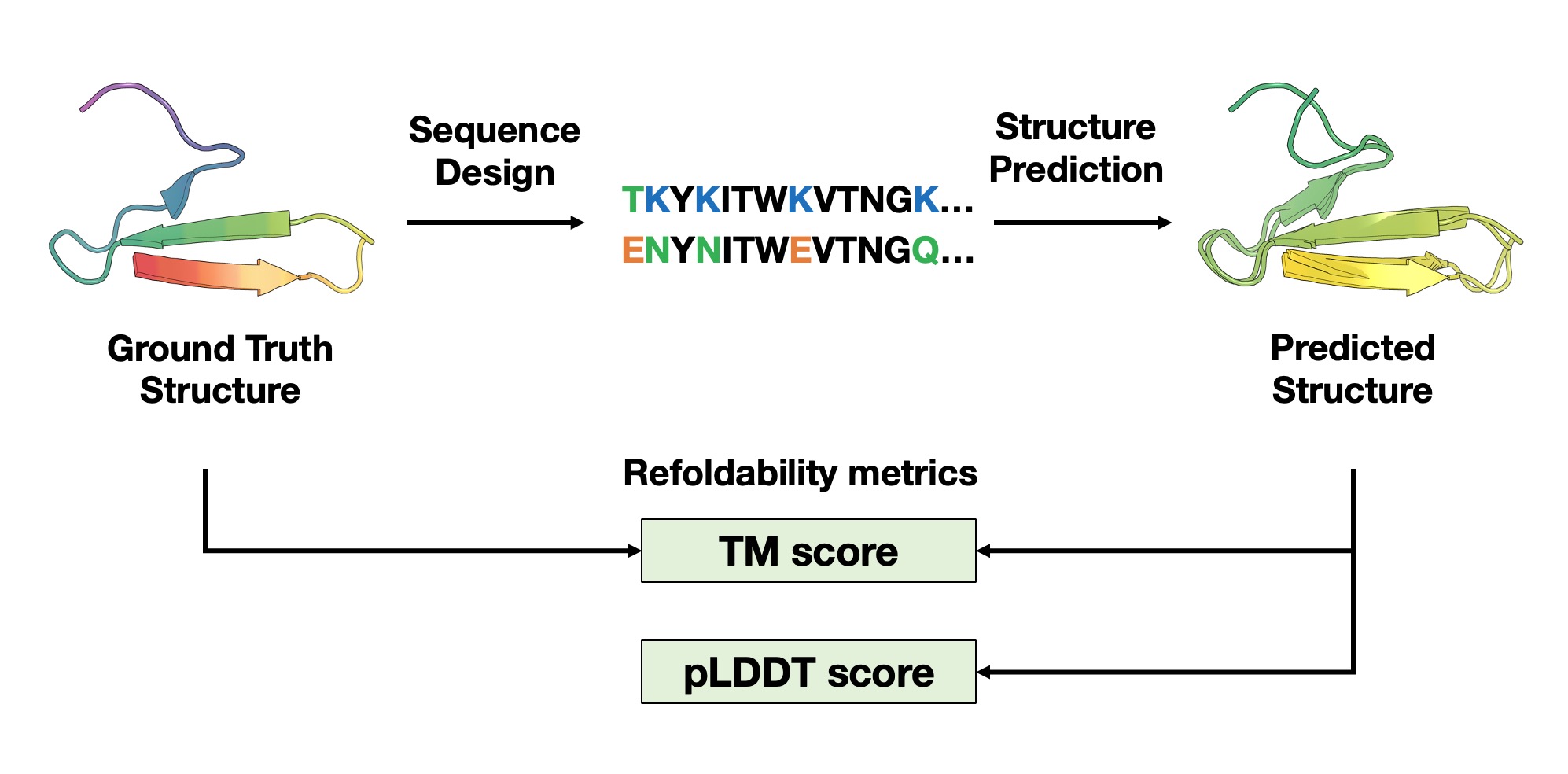}\
\caption{
    Pipeline of measuring refoldability metric. 
}\label{fig:method-refold}
\end{figure}

\section{Method}
\subsection{Evaluating the Designed Sequences with Refoldability-based Metrics}

\paragraph{Motivation}
"Refoldability" is the natural metric that measures the quality of sequences designed based on structures.
It evaluates the sequence quality on two aspects: whether the designed sequence can be expressed and fold stably, and whether they can refold into the input structure.
Previous works have synthesized the proteins experimentally to assess refoldability~\citep{dauparas2022robust, verkuil2022language}.
However, the synthesis process~\citep{mori1999size}, as well as the structure determination methods, such as X-ray crystallography and cryoEM~\citep{ladd1977structure, bai2015cryo} are costly, hindering benchmarking across various design models.
Previously, sequence recovery was proposed as an in silico benchmark~\citep{ingraham2019generative}. While it's straightforward to calculate, there's no confirmed evidence that a high sequence similarity sufficiently implies a high similarity between folded structures, or implies a good foldability in real world.
For instance, even single mutations can cause a protein to misfold, leading to diseases such as Alzheimer's and cystic fibrosis~\citep{cohen2003therapeutic,qu1997cystic}. 
Fortunately, due to advancements in high-accuracy protein structure prediction models, recent work~\citep{wang2022scaffolding} suggests leveraging them as an \emph{in silico} proxy for actual structures.
Adopting this idea, we propose to estimate the true refoldability with structure prediction models.

\paragraph{Metric}
The evaluation pipeline is shown in Figure.~\ref{fig:method-refold}, where we generate multiple sequences with sequence design models given an input structure, and predict the structures for all the generated sequences. Firstly, to assess whether the generated sequences can respect the structure condition, we evaluate the agreement of the ground truth structure with the predicted structures using the TM-score~\citep{zhang2005tm}. We refer this metric as \textbf{Ref-TM}.
Furthurmore, to evaluate the folding stability of the generated sequences, we compute the mean value of the per-residue confidence estimate pLDDT predicted by the structure prediction models, refered as \textbf{Ref-pLDDT}.
Previous research indicates that pLDDT serves as a reliable predictor of disorder~\citep{tunyasuvunakool2021highly}. 
We employ AlphaFold2~\citep{jumper2021highly}, OmegaFold~\citep{wu2022high}, and ESMFold~\citep{lin2023evolutionary} as structure prediction models, which helps minimize deviations due to the choice of model.

It's important to highlight that, although Ref-TM metric and ScTM metric~\citep{trippe2022diffusion} share a similar pipeline, they serve different purposes.
The purpose of ScTM is to evaluate the quality of generated protein structures, treating both the protein design model and structure prediction model as oracles. In contrast, the foldability metric considers the inverse folding model as variable, while maintaining the input structure as a fixed ground truth derived from the test set.




\paragraph{Dataset} 
We use the CATH4.2 40\% non-redundant protein dataset~\citep{orengo1997cath}, and adopt the same data splitting based on CATH topology as StructTrans~\citep{ingraham2019generative}. This results in 18024 protein single chains in the training set, 608 in the validation set, and 1120 in the test set.
We furthur curated a small, high-quality test set from the original test set.
After removing data points with unmeasured coordinates in the protein sequences, we randomly select one protein data from each CATH family and manually excluded proteins with extensive disordered regions,
resulting in a final test set of 82 samples, with length ranges from 49 to 480 amino acids.

\subsection{Evaluating the Estimated Likelihoods with Stability-based Metrics}

\paragraph{Motivation}  
Previous benchmark use perplexity as metric, which is the exponential of negative pseudo-log-likelihood.
However, using and comparing perplexities introduces ambiguity due to several factors.
First, the perplexity value is sensitive to changes in the sampling temperature. Using a protein design method with a high sampling temperature of 0.1, for example, could result in its perplexity exceeding that of a random sampling model based on residue frequency matrix, as demonstrated in Table.~\ref{table-perplexity-value}. 
Second, the computation of pseudo-log-likelihood differs among models, as shown in Table.~\ref{table-perplexity}. For example, PiFold assumes conditional independence of the residue types given the input structure, whereas ESM-IF does not make this assumption. Direct comparison between these methods, therefore, may not be entirely fair.
Lastly, assigning high perplexity to the ground truth sequence does not imply that the protein design method construct the sequence distribution wrongly, since it is possible that the method has distributed a high probability mass function across many sequences which could fold into the given structure but are not present in the dataset.


\begin{minipage}[l]{0.40\textwidth}
  \centering
      \captionof{table}{Perplexities on CATH test set.
      }
  \label{table-perplexity-value}
  \begin{adjustbox}{width=0.98\textwidth}
  \begin{tabular}{ll}
    \toprule
    Design method     & Perplexity    \\
    \midrule
    Uniform & 20.00    \\
    Natural frequencies$^*$     & 18.32     \\
    ESM-IF$(\tau=1)$ & 4.24   \\
    ESM-IF$(\tau=0.1)$     & 3749.51       \\
    \bottomrule
  \end{tabular}
    \end{adjustbox}
\end{minipage}
\begin{minipage}[l]{0.55\textwidth}
\centering
    \captionof{table}{Calculation of pseudo-log-likelihood.
    }
  \label{table-perplexity}
  \begin{adjustbox}{width=0.98\textwidth}
  \begin{tabular}{lc}
    \toprule
    Model Type\footnotemark     & Pseudo-log-likelihood $\gL(\gS \vert \gX, \theta)$\\
    \midrule
    Autoregressive & $ \frac{1}{N} \sum_{i=1}^{N}\log p_{\theta}(s_i \vert s_{<i}, \gX)$    \\
    One-Shot   & $\frac{1}{N} \sum_{i=1}^{N}\log p_{\theta}(s_i \vert \gX)$      \\
    Refinement   & $\frac{1}{N} \sum_{i=1}^{N}\log p_{\theta}(s_i \vert s_{-i}, \gX)$      \\
    MCMC       & $-\lambda_p E_{projection}(\gX \vert \gS) - \lambda_{LM} E_{LM}(\gS)$ \\
    Gradient Descent  & $-\lambda_p E_{projection}(\gX \vert \gS) - \lambda_{LM} E_{LM}(\gS)$ \\
    \bottomrule
  \end{tabular}
  \end{adjustbox}
\end{minipage}

\footnotetext{
The autoregressive decoding models include StructTrans, GVP, ProteinMPNN, and ESM-IF. The one-shot decoding model is represented by PiFold and the refinement decoding model is represented by ByProt. While ESM-Design is categorized under MCMC sampling models, AF-Design is a gradient descent-based model with $\lambda_{LM}=0$.
}

\paragraph{Metric}
Therefore, we propose to use "stability-based metric", which measure whether our structure-based design method can assign higher likelihood to the sequences with higher experimental stablity score.
The score $\gR$ is measured by:
\begin{equation}
    \gR(\theta,\mathcal{D}) = \rho_s\bigl(\gL(S^{(i)} \vert X_{template}, \theta), \gG^{(i)}\bigr)
\end{equation}
where $\rho_s$ is Spearman's correlation, $\theta$ is the design model, $\gL$ is the pseudo-log-likelihood function and $\mathcal{D} = \left\{\gX_{template}, \gS^{(i)}, \gG^{(i)} \right\}$ is the evaluation dataset, with $\gX_{template}$ the template structure, $\gS^{(i)}$ the i-th sequence and $\gG^{(i)}$ the stability score corresponding to the i-th sequence.
If the score $\gR$ is high, the protein design method is likely to assign higher probability to the sequences with higher stability.
Addressing the previously mentioned limitations of the perplexity metric, this dataset with multiple sequences can construct a more accurate sequence landscape that approximate the ground truth distribution.
By the way, we apply the Spearman's correlation to calculate $\gR$, which measure the correlation between score rankings instead of the direct relationship between two attributs, making the protein design methods comparable among them again. Note that the temperature is set to 1.0 for all the models.

\begin{figure}[ht]
    \centering
    \begin{subfigure}[b]{0.49\textwidth}
        \centering
        \includegraphics[width=\textwidth]{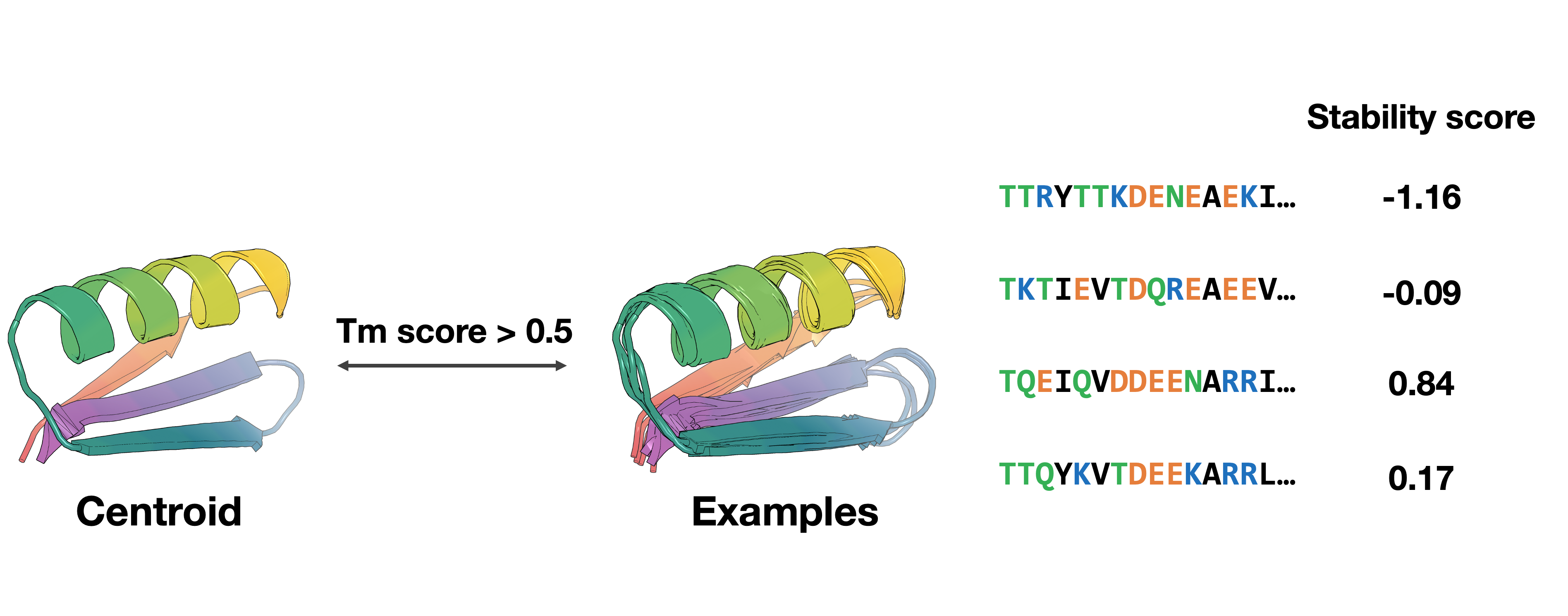}
        \caption{}
        \label{fig:subfig_a}
    \end{subfigure}
    \hfill 
    \begin{subfigure}[b]{0.49\textwidth}
        \centering
        \includegraphics[width=\textwidth]{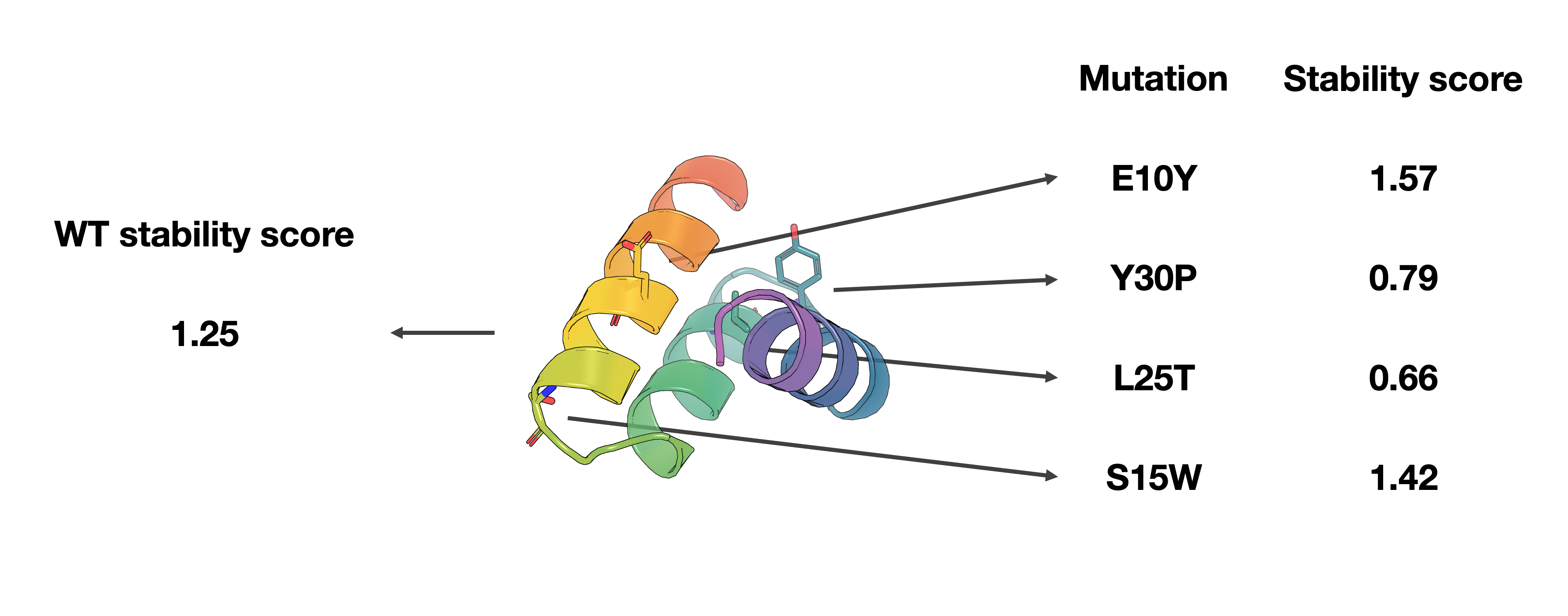}
        \caption{}
        \label{fig:subfig_b}
    \end{subfigure}
    \caption{
    (a) \textit{De Novo} Design dataset with one structure template and four corresponding sequences along with stability scores. 
    (b) Mutagenesis dataset with one structure template and four corresponding sequences along with stability scores.
    }
    \label{fig:method-refoldability}
\end{figure}

\paragraph{Dataset}
We construct such datasets from two types of high-throughput data: "\textit{De Novo} Design"~\citep{rocklin2017global} and "Mutagenesis"~\citep{rocklin2017global, tsuboyama2023mega}. 
The statistics for these datasets are shown in Table.\ref{tab: method_statistic_DeNovo}, Table.\ref{tab:method_table_statistic_Mutagenesis}, and Table.\ref{tab:method_statistic_MutationRocklin} while Figure.\ref{fig:subfig_a} and Figure.~\ref{fig:subfig_b} illustrate example datasets.
The first category, \textit{De Novo} Design data, refers to proteins modeled after specific structural templates. These proteins are designed based on these structure templates, and they will fold into corresponding structure once it can be folded. 
Even though there is a one-to-one relationship between structure and sequence in this data, structures stemming from the same topology show only subtle differences.
For our curated dataset, we clustered these structural templates and replaced individual templates with the centroid of their respective clusters.
Given that all structures within a cluster have a TMscore exceeding 0.5 with each other, it is reasonable to assume that sequences derived from these structures would have highly similar folds~\citep{xu2010significant}.
In contrast, the second category, Mutagenesis data, is derived from various templates, including both natural proteins in the PDB and \textit{De Novo} designed proteins with predicted structures. These datasets contain a significant amount of single-site and \revise{double-site} mutation data related to the corresponding template, providing insight into which mutations stabilize the protein.\footnote{The stability-based metric evaluated on Mutagenesis dataset is similar to the experiment conducted by \cite{ingraham2019generative}. However, while \cite{ingraham2019generative} applied the Pearson correlation score.} 
\revise{We further removed the 'insertion' and 'deletion' types of mutations, which alter the length of amino acid sequences, from the original dataset~\citep{tsuboyama2023mega}. This resulted in 527K sequences with a stability score.}

\begin{table*}[ht]
    \centering
    \caption{Dataset statistic of the \textit{De Novo} Design data. 
    We clustered the protein structures and selected the four largest clusters as datasets based on two design topologies: $EHEE$ and $HHH$. For instance, $EHEE_6$ denotes the 6th structural cluster, which is the largest cluster within the $EHEE$ topology. Sequences are considered stable if their experimental stability score is greater than or equal to 1.
    We are working on the structural clustering of \textit{de novo} designed proteins to curate more datasets.
    }
    \label{tab: method_statistic_DeNovo}
    \begin{adjustbox}{max width=\linewidth}
        \begin{tabular}{lllllllll}
        \toprule
        \textbf{} & $EHEE_3$ & $EHEE_4$ & $EHEE_5$ & $EHEE_6$ & $HHH_{54}$ & $HHH_{82}$ & $HHH_{84}$ & $HHH_{86}$ \\ \midrule
        \textbf{\# sequence} & 1743 & 1850 & 477 & 6873 & 669 & 632 & 1990 & 612 \\ 
        \textbf{\# stable sequence} & 110 & 120 & 31 & 511 & 203 & 186 & 621 & 213 \\ 
        \textbf{portion} & 0.06 & 0.06 & 0.06 & 0.07 & 0.30 & 0.29 & 0.31 & 0.35 \\ \bottomrule
    \end{tabular}
    \end{adjustbox}
\end{table*}

\begin{table}[!ht]
    \centering
    \caption{Dataset statistic of the Mutagenesis data derived from \cite{rocklin2017global}. Sequences are considered stable if their experimental stability score is greater than or equal to 1}
    \label{tab:method_table_statistic_Mutagenesis}
    \begin{adjustbox}{width=\textwidth}
        \begin{tabular}{llllll}
        \hline
            \textbf{} & $EEHEE_{37}$ & $EEHEE_{1498}$ & $EEHEE_{1702}$ & $EEHEE_{1716}$ & $EEHEE_{779}$ \\ \hline
            \textbf{\# sequence} & 775 & 775 & 775 & 775 & 775 \\ 
            \textbf{\# stable sequence} & 49 & 392 & 680 & 339 & 163 \\ 
            \textbf{portion} & 0.063 & 0.506 & 0.877 & 0.437 & 0.21 \\ \hline
        \end{tabular}
    \end{adjustbox}
    \begin{adjustbox}{width=\textwidth}
    \begin{tabular}{lllllll}
    \hline
        \textbf{} & $HEEH_{223}$  & $HEEH_{726}$  & $HEEH_{872}$ & $HHH_{142}$ & $HHH_{134}$ & $HHH_{138}$ \\ \hline
        \textbf{\# sequence} & 775 & 775 & 775 & 775 & 775 & 775 \\ 
        \textbf{\# stable sequence} & 453 & 39 & 438 & 623 & 720 & 754 \\ 
        \textbf{portion} & 0.585 & 0.05 & 0.565 & 0.804 & 0.929 & 0.973 \\ \hline
    \end{tabular}
    \end{adjustbox}
\end{table}

\begin{table}[!ht]
    \centering
    \caption{Dataset statistic of the Mutagenesis data derived from dataset \#3 in \cite{tsuboyama2023mega}}
    \label{tab:method_statistic_MutationRocklin}
    \begin{adjustbox}{width=\textwidth}
    \begin{tabular}{lccccc}
    \toprule
    Dataset                    & Description                                             & \# of total sequences & sequence group       & \# sequences groups & \# of sequences \\
    \midrule
            \multirow{2}{*}{Original Dataset} &
            All data for $\Delta \Delta G$ &
            \multirow{2}{*}{607,839}&
            single-site mutation&412~wild-types&448,788 \\
            \cmidrule{4-6}
            & (WT < 4.75 kcal/mol) && double-site mutation & 496 pairs & 159,051 \\
        \midrule
            \multirow{2}{*}{Filtered Dataset } &
            \multirow{2}{*}{remove "indel" and "delete"} &
            \multirow{2}{*}{527,830}&
            single-site mutation& 372 wild-types& 368,779 \\
            \cmidrule{4-6}
            & && double-site mutation & 481 pairs & 159,051 \\              
    \bottomrule
    \end{tabular}
    \end{adjustbox}
\end{table}


\section{Experiments}
\label{gen_inst}
\paragraph{Baselines}
We evaluate StructTrans~\citep{ingraham2019generative}, GVP~\citep{jing2020learning}, ProteinMPNN~\citep{dauparas2022robust}, PiFold~\citep{gao2023pifold}, ByProt~\citep{zheng2023structure}, AF-Design\footnote{https://github.com/sokrypton/ColabDesign}~\citep{wang2022scaffolding}, ESM-Design~\citep{verkuil2022language}, ESM-IF1~\citep{hsu2022learning} with our benchmark. 
Each model follows the default settings provided in their original papers or codebases. 
Encoder-decoder models were originally trained on the CATH4.2 train dataset for up to 100 epochs. 
ESM-IF1, on the other hand, was originally trained on the CATH4.3 data set\footnote{Since ESM-IF1 is trained on CATH4.3, we did not evaluate its refoldability on CATH4.2 to avoid potential data leakage. Currently, we cannot train ESM-IF on CATH4.2 because the training code has not been provided.}. 
ESM-Design and AF-Design models were trained on full UniRef data~\citep{suzek2015uniref}or complete PDB data~\citep{jumper2021highly, berman2000protein}.
We chose to overlook potential data leakage issues for these models because they begin from random starting points and is inable to sample the exact ground truth sequence accurately, as demonstrated in further experiments.
All experiments were conducted on Nvidia Quadro RTX8000.


    

\subsection{Benchmarking on Refoldability-based Metrics}
\paragraph{Settings} For each structure in the test set, we randomly generated 100 sequences using protein design models and also from random models that sample from uniform and natural frequency distributions.
The sampling temperature is set to 0.1 for all encoder-decoder models. 
However, due to the slow inference speed of AF-Design (2.4 GPU hours per sequence on average) and ESM-Design (9 GPU hours per sequence on average), we limited our generation to 5 sequences per structure for AF-Design and just 1 sequence per structure for ESM-Design\footnote{would the results be statistical significant? Also this 1 vs. 5, would this cause bias? → Elaborate that, (1) this methods are too slow, we cannot afford training it, (2) with 1 sample generated, we can already have a coarse-grained observation on the performance, and see that their performance is largerly lagged behind other methods.}.
We then predicted the structures of these generated sequences using both ESMFold and OmegaFold.
AlphaFold2\footnote{We use the ColabFold~\citep{mirdita2022colabfold} implementation with MMseqs MSA alignment~\citep{steinegger2017mmseqs2, mirdita2019mmseqs2}.} is somehow time costly to run, so we randomly feed one sequence per structure into AlphaFold2.
Finally, we employ the TMalign toolkit~\citep{zhang2005tm} to compute the Ref-TM score.

\begin{table}[ht]
\centering
\caption{Refoldability metric and recovery metric on the CATH dataset. We employ \textbf{bold} and \underline{underlining} to highlight the best and suboptimal results on each metric. We use TM and pLDDT to represent Ref-TM and Ref-pLDDT.}
\label{tab:exp-refold-main}
\begin{adjustbox}{width=\textwidth}
        \begin{tabular}{lcccccccccc}
            \toprule
            \multirow{2}{*}{\textbf{Design method}} &
            \multicolumn{2}{c}{ESMFold}&&
            \multicolumn{2}{c}{OmegaFold}&& \multicolumn{2}{c}{AlphaFold2}
            && \multirow{2}{*}{Recovery\%}\\
            \cmidrule{2-3}
            \cmidrule{5-6}
            \cmidrule{8-9}

        & TM & pLDDT && TM & pLDDT&& TM & pLDDT && \\ \midrule
\textbf{Uniform}             & 0.05    & 27.68 && 0.05      & 31.53 && 0.06 & 33.68 && 5.00           \\
\textbf{Natural frequencies} & 0.07    & 30.53 && 0.07      & 35.59 && 0.06 & 35.02 && 5.84              \\ \midrule
\textbf{StructTrans}         & 0.72    & 68.85 && 0.64      & 70.35 && 0.79 & 80.66 && 35.89              \\
\textbf{GVP}                 & 0.73    & 69.67 && 0.67      & 74.33 && 0.83 & 84.29 && 39.46              \\
\textbf{ProteinMPNN}         & \textbf{0.80}    & \textbf{76.53} && \textbf{0.76}      & \textbf{80.75} && \textbf{0.87} & \textbf{87.89} && 41.44             \\
\textbf{PiFold}              & 0.71    & 67.55 && 0.64      & 70.21 && 0.82 & 82.54 && \underline{44.86}              \\
\textbf{ByProt}              & \underline{0.73}    & \underline{72.12} && \underline{0.70}      & \underline{77.58} && \underline{0.85} & \underline{87.26} && \textbf{51.23}             \\ \midrule
\textbf{AF-Design}           & 0.53    & 61.37 && 0.53      & 72.04 && 0.52 & 75.29 && 15.95           \\ 
\textbf{ESM-Design}          & 0.38    & 59.65      && 0.38          &62.66       &&0.37      &60.02    && 17.33                \\ \midrule
\textbf{Wildtype}                  & 0.80    & 74.91 && 0.75      & 78.39 && 0.90 & 89.87 && 100                 \\ \bottomrule
\end{tabular}
\end{adjustbox}
\end{table}

\paragraph{Refoldability Metric Analysis}
We report the refoldability and recovery metrics in Table.~\ref{tab:exp-refold-main}. We observe that ProteinMPNN stands out as the leading design method across the refoldability metrics, attaining 0.87 Ref-TM and 87.89 Ref-pLDDT with AlphaFold2 prediction. 
ByProt is slightly behind with a 0.85 Ref-TM and 87.26 Ref-pLDDT, followed by GVP.
The sampling-based model ESM-Design and gradient-based model AF-Design are subpar in terms of both recovery and refoldability. 
We note that despite low recovery, the structure somewhat resembles the input structure. 

\paragraph{Refoldability Metric on Ground Truth Sequence and Random Sequences}
To provide a context, we tested the refoldability of sequences generated by "non-learnable models" and wildtype sequences. 
(i) Sequences generated by random sampling have very low refoldability metric. 
Their Ref-TM is only around 0.05, suggesting a significant dissimilarity between the folded structure and the input structure.
Additionally, the low pLDDT value implies poor sequence quality, indicating difficulty in proper folding. 
(ii) In contrast, for wildtype sequences, the AlphaFold2 model predicts a higher Ref-TM of 0.90 and Ref-pLDDT of 89.87. This suggests that we can trust AlphaFold's predictive accuracy for novel sequences.

\paragraph{Discord between Recovery and Refoldability}
Table.~\ref{tab:exp-refold-main} also indicates that the recovery metric and refoldability metrics are not fully aligned. 
For example, 
(i) ProteinMPNN, which ranks 3rd in the recovery metric with a 41.44\% recovery, lags behind ByProt's 51.23\% recovery. However, ProteinMPNN achieves the highest scores on both Ref-TM and Ref-pLDDT metrics;
(ii) PiFold, despite having the second-highest recovery at 44.86\%, only ranks fourth in refoldability metrics, lagging behind ByProt, GVP, and ProteinMPNN.

\paragraph{Consistency between Structure Prediction Models} 
We noticed that the ranking of different protein design methods remains consistent when we apply different structure prediction models.
This further bolsters the credibility of the refoldability metrics. 
Moreover, if a protein design method excels under ESMFold, it will likely perform similarly well under OmegaFold and AlphaFold2.

\paragraph{Consistency between Ref-TM and Ref-pLDDT}
We find that the trends for Ref-TM and Ref-pLDDT computed based on the three structure prediction models are strikingly similar. 
When the Ref-pLDDT is higher, the Ref-TM is typically higher as well.
Consequently, for structure-based protein design, we recommend to use structural prediction models as discriminators to pre-screen the generated sequences with Ref-pLDDT.



\subsection{Benchmarking on Stability-based Metrics}

\paragraph{Result Anaylsis on \textit{De Novo} Design Data} Table.~\ref{exp_table_denovo} shows the stability metric on \textit{De Novo} Design datasets. We observe that (i) AF-Design exhibits the highest correlation with stability scores, likely due to the utilization of AlphaFold2. However, sampling from the estimated distribution remains a challenge; (ii) Within the encoder-decoder family, ESMIF performs the best, followed by ByProt and ProteinMPNN; 
(iii) Surprisingly, ESM-Design does not perform as good as AF-Design model, and also falling short compared to other encoder-decoder methods.

\begin{table}[!ht]
    \centering
    \caption{Stability metric on \textit{De Novo} Design datasets.
    }
    \label{exp_table_denovo}
    \begin{adjustbox}{width=\textwidth}
    \begin{tabular}{lccccccccc}
    \toprule
        \textbf{Design method}  & $EHEE_3$ & $EHEE_4$ & $EHEE_5$ & $EHEE_6$ & $HHH_{54}$ & $HHH_{82}$ & $HHH_{84}$ & $HHH_{86}$ & \textbf{mean} \\ \midrule
        \textbf{GVP}  & 0.158 & 0.285 & 0.299 & 0.271 & 0.682 & 0.593 & 0.588 & 0.658 & 0.442 \\ 
        \textbf{PiFold} & 0.158 & 0.287 & 0.269 & 0.267 & \underline{0.688} & 0.607 & 0.556 & 0.641 & 0.434\\ 
        \textbf{ProteinMPNN}  & 0.176 & 0.282 & 0.314 & 0.274 & \underline{0.688} & 0.584 & 0.570 & 0.626 & 0.439\\ 
        \textbf{ESMIF}  & 0.171 & \underline{0.335} & \underline{0.331} & \underline{0.282} & 0.678 & \underline{0.660} &  \underline{0.625} & \underline{0.691} &\underline{0.472}\\ 
        \textbf{ByProt}  & \underline{0.191} & 0.297 & 0.296 & 0.270 & \underline{0.688} & 0.631 & 0.571 & 0.626 & 0.446\\ \midrule
        \textbf{AF-Design}  & \textbf{0.252} & \textbf{0.366} & \textbf{0.402} & \textbf{0.353} & \textbf{0.699} & \textbf{0.672} & \textbf{0.661} & \textbf{0.723} & \textbf{0.516}\\ 
        \textbf{ESM-Design}  & 0.153 & 0.259 & 0.291 & 0.189 & 0.622 & 0.369 & 0.303 & 0.362 & 0.319\\  \bottomrule
    \end{tabular}
    \end{adjustbox}
\end{table}

\paragraph{Result Anaylsis on Mutagenesis Data}
Table.~\ref{exp_table_SSM} shows the stability metric on Mutagenesis datasets presented in \citep{rocklin2017global}. (i) There is no single model that consistently performs well across all datasets. Overall, ESM-IF again achieves the highest mean correlation score of 0.433, and PiFold achieves the second with 0.413 correlation. 
(ii)  The observation that PiFold performs well in density estimation on the mutational dataset and in recovery suggests that PiFold excels at modeling per-residue likelihood.
(iii) The performance of AF-Design and ESM-Design is subpar. The possible reason is that structure prediction based models are not sensitive to point mutations~\citep{pak2023using}.

\begin{table}[!ht]
    \centering
    \caption{
    Stability metric on Mutagenesis datasets in \citep{rocklin2017global}.
    }
    \label{exp_table_SSM}
    \begin{adjustbox}{width=\textwidth}
    \begin{tabular}{lcccccc}
    \toprule
                \textbf{Design method} & $EEHEE_{37}$ & $EEHEE_{1498}$ & $EEHEE_{1702}$ & $EEHEE_{1716}$ & $EEHEE_{779}$ & $HEEH_{223}$\\ \midrule
        \textbf{GVP} & 0.481 & 0.318 & \underline{0.247} & 0.413 & 0.526 & 0.340 \\ 
        \textbf{PiFold} & 0.581 & 0.298 & 0.187 & 0.477 & 0.580  & \underline{0.413}\\ 
        \textbf{ProteinMPNN} & 0.597 & \underline{0.382} & 0.136 & 0.384 & \underline{0.595} & 0.324 \\ 
        \textbf{ESMIF} & \textbf{0.641} & \underline{0.382} & 0.236 & \textbf{0.565} & \textbf{0.645} & \textbf{0.454} \\ 
        \textbf{ByProt} & \underline{0.629} & \textbf{0.414} & \textbf{0.320} & \underline{0.548} & 0.584  & 0.402 \\  \midrule
        \textbf{AF-Design} & 0.557 & 0.300 & 0.027 & 0.036 & 0.490 & 0.195  \\ 
        \textbf{ESM-Design} & 0.240 & 0.115 & -0.080 & 0.188 & 0.039 & 0.227\\ \bottomrule
    \end{tabular}
    \end{adjustbox}
    \begin{adjustbox}{width=0.8\textwidth}
    \begin{tabular}{lcccccc}
    \toprule
        \textbf{Design method}   & $HEEH_{726}$  & $HEEH_{872}$ & $HHH_{142}$ & $HHH_{134}$ & $HHH_{138}$ & \textbf{mean}\\ \midrule
        \textbf{GVP}  & 0.102 & 0.248 & 0.502 & 0.253 & 0.295 & 0.339\\ 
        \textbf{PiFold}  & \textbf{0.239} & 0.315 & \underline{0.536} & 0.290 & \underline{0.383} &\underline{0.391}\\ 
        \textbf{ProteinMPNN}  & -0.055 & 0.205 & 0.431 & 0.256 & 0.326 &0.326\\ 
        \textbf{ESMIF}  & 0.216 & \underline{0.335} & \textbf{0.573} & \underline{0.318} & \textbf{0.398} &\textbf{0.433}\\ 
        \textbf{ByProt} & \underline{0.238} & \textbf{0.338} & 0.511 & 0.289 & 0.360 &0.421\\ \midrule
        \textbf{AF-Design}  & 0.214 & -0.148 & 0.453 & \textbf{0.351} & 0.314 &0.254\\ 
        \textbf{ESM-Design}  & 0.062 & 0.013 & 0.004 & -0.050 & -0.050 &0.064\\ \bottomrule
    \end{tabular}
    \end{adjustbox}
\end{table}

\paragraph{Result Anaylsis on Mega-Scale Mutagenesis Data} 
Table~\ref{exp_table_RocklinMutation} presents the stability metric applied to mega-scale Mutagenesis datasets, which includes 527,830 sequences. This dataset is significantly larger than the one comprising 8,525 sequences used in the previous table. We have divided the dataset into two parts: mutations on 215 natural proteins and mutations on 156 \textit{de novo} designed proteins. This division allows us to examine whether the models perform differently on these groups. The correlation scores were first calculated for each sequence group relative to its corresponding wild-type protein, and then these scores were averaged. Our observations are as follows: (i) ESMIF consistently achieves the highest correlation scores across both \textit{de novo} proteins and natural proteins, followed by ProteinMPNN, ByProt, and PiFold; (ii) Encoder-decoder based models show lower correlation scores on \textit{de novo} sequence groups, while structure-prediction based models attain higher scores on natural sequence groups;
(iii) The performance of AF-Design and ESM-Design remains subpar in this larger dataset. Notably, ESM-Design performs poorly on natural proteins, exhibiting both positive and negative Spearman's correlation, which results in an average correlation score near zero.

\begin{table}[!ht]
    \centering
    \caption{Stability metric applied on mega-scale experimental Mutagenesis datasets~\citep{tsuboyama2023mega}. The columns display the average stability scores for \textit{de novo} designed proteins, natural proteins in the PDB, and across all 372 sequence groups.}
    \label{exp_table_RocklinMutation}
    \begin{tabular}{lccc}
    \toprule
        \textbf{Design method} & \textbf{\textit{De Novo}} & \textbf{Natural} & \textbf{All} \\
    \midrule
        \textbf{GVP}	&0.390	&0.494  &0.450\\  
        \textbf{PiFold}    	&0.448	&0.556&0.511\\  
        \textbf{ProteinMPNN}    	&0.428	&\underline{0.605}&0.531\\  
        \textbf{ESMIF}    	&\textbf{0.500}	&\textbf{0.629}&\textbf{0.575}\\  
        \textbf{ByProt}    	&\underline{0.468}	&0.586&\underline{0.536}\\  
        \midrule
        \textbf{AF-Design}    	 &0.354	&0.292&0.318\\  
        \textbf{ESM-Design}   	&0.127	&0.0004 &0.053\\  

    \bottomrule
    \end{tabular}
\end{table}





\subsection{Takeaways}
\begin{itemize}
    \item 
    Our findings indicate differences between the recovery metric and the refoldability metrics introduced in our benchmark. Notably, a model with high recovery doesn't necessarily guarantee good refoldability.
    \item 
    Results suggest that encoder-decoder methods generally outperform structure-prediction based methods in terms of refoldability, recovery, and stability metrics. However, the structure-prediction based methods show potential in accurate sequence density estimation, which may lead to the generation of superior sequences.
    \item 
    Among encoder-decoder methods, ByProt, ProteinMPNN, and ESM-IF show strong performance on our benchmark. We observe that PiFold excels in recovery metrics, but its performance in refoldability and stability metrics is less impressive. This may potentially be related to the conditional independence assumption applied during sampling and density estimation.
    \item 
    Among structure-prediction based methods, AF-Design method has advantages over the ESM-Design method in various metrics, also including inference efficiency.
\end{itemize}

\section{Conclusion and Future Work}

\paragraph{Conclusion}
To better evaluate structure-based protein design models, we propose the refoldability-based metric and stability-based metric. We curate datasets corresponding to these metrics, and conduct experiments on this \textbf{PDB-Struct} benchmark.
By examining the benchmark results, we detail the strengths and weaknesses of each type of protein design model, offering insights to protein researchers in their model selection. This paves the way for a fair and comprehensive evaluation of protein design methods in the future.

\paragraph{Future work}
We are continuously collecting additional \textit{De Novo} Design and Mutagenesis datasets to enhance our benchmark, and we are evaluating newly released protein-design methods such as KW-design~\citep{gao2023knowledge} and GRADE-IF~\citep{yi2023graph}.
Furthermore, we are conducting extensive experiments to demonstrate the superiority of refoldability metrics over the recovery metric.
Discussions regarding the efficiency and reliability of the \textbf{PDB-Struct} benchmark evaluations are ongoing, and we intend to address these in a future version of this work.
Concurrently, we discovered another project for benchmarking protein design methods, ProteinInvBench~\citep{gao2023proteininvbench}, which has been accepted into the NeurIPS 2023 Datasets and Benchmarks Track. Inspired by their approach, we are considering the addition of a diversity metric to our benchmark.

\clearpage
\section*{Acknowledgement}
We would like to thank to Zuobai Zhang, Yangtian Zhang for their constructive feedback on the manuscript. This project is supported by Twitter, Intel, the
Natural Sciences and Engineering Research Council (NSERC) Discovery Grant, the Canada CIFAR
AI Chair Program, Samsung Electronics Co., Ltd., Amazon Faculty Research Award, Tencent AI Lab
Rhino-Bird Gift Fund, a NRC Collaborative R\&D Project (AI4D-CORE-06) as well as the IVADO
Fundamental Research Project grant PRF-2019-3583139727.

\bibliographystyle{plainnat} 
\bibliography{Benchmark}


\end{document}